\documentclass[showpacs,amsmath,amssymb,aps,longbibliography,twocolumn,pra]{revtex4-2}

\usepackage[utf8]{inputenc}

\usepackage{graphicx}

\usepackage{etoolbox}
\makeatletter
\pretocmd{\NAT@open}{\begingroup\color{\@citecolor}}{}{}
\apptocmd{\NAT@close}{\endgroup}{}{}
\makeatother

\newenvironment{methods}%
  {\acknowledgments}%
  {\endacknowledgments}

\usepackage{xr}
\usepackage{braket}
\usepackage{bbm}

\usepackage{placeins}
\usepackage{soul}
\usepackage[usenames]{color}
\definecolor{dark-blue}{rgb}{0,0.2,0.6}

\usepackage[english]{babel}
\usepackage{enumerate}
\usepackage{pifont}
\usepackage{amsfonts}
\usepackage{siunitx}
\usepackage{gensymb}

\usepackage{amssymb}
\usepackage{wasysym}
\usepackage{physics}
\usepackage{bbold}

\usepackage{amsmath}
\usepackage{mathtools}

\usepackage{upgreek}

\usepackage{xspace}
\usepackage{titletoc}

\usepackage[colorlinks,%
            pdfusetitle,%
            urlcolor=dark-blue,%
            citecolor=dark-blue,%
            linkcolor=dark-blue]{hyperref}

\newcommand{\be}{\begin{equation}}
\newcommand{\ee}{\end{equation}}

\newcommand{\subfigref}[2]{\hyperref[fig:#1]{\ref*{fig:#1}(#2)}}

\renewcommand{\eqref}[1]{Eq.\,\ref{#1}}

\DeclareMathOperator{\Imag}{Im}

\begin{document}

\title{Probing disorder-driven topological phase transitions via topological edge modes \\ with ultracold atoms in Floquet-engineered honeycomb lattices}

\author{Alexander~Hesse}
\author{Johannes~Arceri}
\author{Moritz~Hornung}
\author{Christoph~Braun}
\author{Monika~Aidelsburger}
\email{monika.aidelsburger@mpq.mpg.de}
\affiliation{Max-Planck-Institut f\"{u}r Quantenoptik, 85748 Garching, Germany}
\affiliation{Munich Center for Quantum Science and Technology (MCQST), 80799 M\"{u}nchen, Germany}
\affiliation{Fakult\"{a}t f\"{u}r Physik, Ludwig-Maximilians-Universit\"{a}t, 80799 M\"{u}nchen, Germany}

\date{\today}

\begin{abstract}
One of the most fascinating properties of topological phases of matter is their robustness to disorder and imperfections. Although several experimental techniques have been developed to probe the geometric properties of engineered topological Bloch bands with cold atoms, they almost exclusively rely on the translational invariance of the underlying lattice. This prevents direct studies of topology in the presence of disorder, further hindering an extension to disordered interacting topological phases. Here, we identify disorder-driven phase transitions between two distinct Floquet topological phases using the characteristic properties of topological edge modes with ultracold atoms in periodically-driven two-dimensional (2D) optical lattices. Our results constitute an important step towards studying the rich interplay between topology and disorder with cold atoms. Moreover, our measurements confirm that disorder indeed favors the anomalous Floquet topological regime over conventional Hall systems, indicating an enhanced robustness and paving the way towards observing exotic out-of-equilibrium phases such as the anomalous Floquet Anderson insulator. 
\end{abstract}

\maketitle

\begin{figure}[!htb]
\centering
\includegraphics{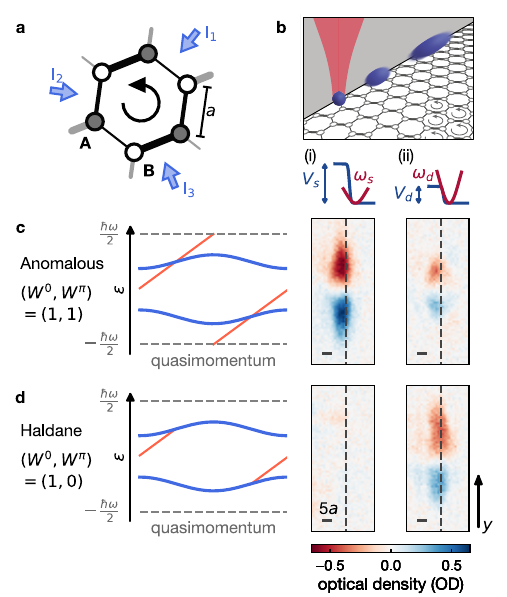}
\caption{\textbf{Experimental setup and measurement protocol.} \textbf{a} Schematic of the periodically-modulated honeycomb lattice generated by interfering three laser beams with variable intensities $I_i(t)$, $i=\{1,2,3\}$ and $a=287\,$\si{\nano\meter}. The width of the tunneling bonds indicates their variable strength, modulated in a chiral manner with frequency $\omega$. \textbf{b} Schematic of the preparation of edge modes at a topological interface: A tightly-focused optical tweezer (red) confines a BEC (blue), which is prepared and released at a repulsive potential step (gray). The time evolution of the initially confined wave packet is illustrated by ellipsoids (blue), propagating along the topological interface. Bottom: Two different experimental settings to tune the initial-state: (i) Shallow tweezer with radial harmonic oscillator frequency $\omega_s/(2\pi) = \SI{1.3\pm0.1}{\kilo\hertz}$ and large potential step $V_s/h =\SI{13.2}{\kilo\hertz}$. (ii) Deep tweezer with radial frequency $\omega_d/(2\pi) = \SI{2.0\pm0.1}{\kilo\hertz}$ and potential height $V_d/h = \SI{2.8}{\kilo\hertz}$. \textbf{c,d} Left: Winding numbers $(W^0,W^\pi)$ and schematic bandstructure (blue) of the \textbf{c} anomalous Floquet [$\omega/(2\pi) = 7\,$\si{\kilo\hertz}] and \textbf{d} Haldane [$\omega/(2\pi) =  16\,$\si{\kilo \hertz}] regime. Edge dispersion is marked in red. Right: Difference of absorption pictures (averaged over 100 images per chirality) obtained for the two different chiralities $\kappa=\pm 1$ after 50$T$, with $T=2\pi/\omega$, for the two experimental settings (i,ii) illustrated at the top. The dashed vertical line indicates the location of the topological interface.
}
\label{fig:Fig1}
\end{figure}

The discovery of topological phases of matter has revolutionized condensed matter physics~\cite{qi_topological_2011,wen_colloquium_2017}. Remarkably, disorder plays a crucial role in revealing topological robustness. A prominent example is the quantization of Hall plateaus in the integer~\cite{von_klitzing_quantized_1986} and fractional quantum Hall effects~\cite{stormer_fractional_1999}. While strong disorder eventually induces a transition to a trivial regime when its amplitude approaches the spectral gap, weak disorder is essential for pinning the Fermi energy within the gap, thereby stabilizing the plateaus with extraordinary precision.

Synthetic quantum systems offer unique opportunities to study the properties and exotic excitations of topological phases of matter~\cite{aidelsburger_artificial_2018,cooper_topological_2019} with microscopic resolution and control~\cite{gross_quantum_2021}. With neutral atoms in optical lattices, topological phases are often engineered via periodic driving (Floquet engineering)~\cite{bukov_universal_2015,eckardt_colloquium_2017,weitenberg_tailoring_2021}, or synthetic dimensions~\cite{celi_synthetic_2014,mancini_observation_2015,livi_synthetic_2016,kolkowitz_spinorbit-coupled_2017}. This has enabled the realization of Hofstadter~\cite{aidelsburger_realization_2013,miyake_realizing_2013,mancini_observation_2015,livi_synthetic_2016,tai_microscopy_2017} and Haldane-type models~\cite{jotzu_experimental_2014,tarnowski_measuring_2019,Wintersperger2020}, with recent advances in studying strongly-interacting regimes~\cite{leonard_realization_2023,zhou_observation_2023,impertro_strongly_2025}.  
Moreover, numerous techniques have been developed to characterize the engineered geometric properties of Floquet-Bloch bands~\cite{cooper_topological_2019} with high precision~\cite{atala_direct_2013,duca_aharonov-bohm_2015,jotzu_experimental_2014,aidelsburger_measuring_2015,flaschner_experimental_2016,tarnowski_measuring_2019,asteria_measuring_2019,braun_real-space_2024}. While this provides an excellent starting point for studying the rich interplay between topology and disorder~\cite{titum_disorder-induced_2015,titum_anomalous_2016,hofstetter_phase_transition,hofstetter_AFAI}, experimental studies remain scarce.

Topological pumps~\cite{citro_thouless_2023} -- 1D dynamical versions of 2D topological systems -- offer a simpler experimental and theoretical route for investigating topology and disorder~\cite{niu_quantised_1984}. Indeed, disorder-induced phase transitions have been studied in the context of 1D topological Thouless pumps in photonic waveguide arrays~\cite{cerjan_thouless_2020}, with neutral atoms~\cite{nakajima_competition_2021}, superconducting qubits~\cite{liu_interplay_2025} and in metamaterials~\cite{grinberg_robust_2020}. Disorder can further drive a transition from a topologically trivial to a non-trivial phase known as the topological Anderson insulator, which has been studied experimentally in 1D topological pumps~\cite{nakajima_competition_2021} and 1D momentum-space lattices~\cite{meier_observation_2018}, while 2D realizations have so far only been reported in photonic systems~\cite{stutzer_photonic_2018,liu_topological_2020,chen_realization_2024}.

Chiral edge modes serve as clear signatures of phase transitions between topologically trivial and non-trivial regimes~\cite{halperin_quantized_1982,hatsugai_chern_1993}.
In this work, we extend this idea by demonstrating that mode matching between a localized Bose-Einstein condensate (BEC) -- precisely controlled via an optical tweezer~\cite{braun_real-space_2024} -- and topologically protected chiral edge modes can be exploited to selectively prepare edge modes in distinct Floquet topological regimes. This enables the study of topological phase transitions in the presence of disorder, which we introduce using an optical speckle potential~\cite{bouyer_quantum_2010,Goodman_speckle}. Specifically, we investigate disorder-driven transitions between the anomalous Floquet (AF) and Haldane (H) topological regimes in a 2D periodically-modulated honeycomb lattice (Fig.~\ref{fig:Fig1}a). Our experimental observations are in agreement with numerical simulations, underscoring the reliability of our approach. Central to this is a newly developed method for calibrating the average disorder strength, based on Kapitza-Dirac diffraction of a BEC~\cite{supplement}. In contrast, earlier methods relied on optical calibrations in a conjugate imaging plane~\cite{clement_suppression_2005} or state-dependent disorder potentials~\cite{volchkov_measurement_2018,rubio-abadal_many-body_2019} limiting their applicability to more general settings.

Our technique for probing topological disordered systems builds on our recent work on the observation of topological chiral edge modes~\cite{braun_real-space_2024}: A localized BEC is confined in a tightly-focused optical tweezer trap near a topological interface generated via a programmable repulsive potential step (Fig.~\ref{fig:Fig1}b). After release, the in-situ evolution of the atoms reveals the presence or absence of chiral motion. Distinct Floquet topological phases are realized via chiral modulation of the three tunnel couplings (Fig.~\ref{fig:Fig1}a), as demonstrated in Ref.~\cite{Wintersperger2020}. Floquet topological phases are characterized by so-called winding numbers, which determine the number and chirality of topological edge modes in a specific quasienergy gap. For a generic two-band model there are two different energy gaps, one around zero energy (0-gap) and one at the Floquet Brillouin zone edge ($\pi$-gap) with winding numbers ($W^0$,$W^\pi$). The generalized bulk-boundary correspondence connects the Chern number of each band to the winding number according to: $\mathcal{C}^{\mp}=\pm(W^0-W^\pi)$~\cite{kitagawa_topological_2010,rudner_anomalous_2013,nathan_topological_2015}, where $\mathcal{C}^{-}$ ($\mathcal{C}^{+}$) denotes the Chern number of the lower (upper) band. Here, we focus on two distinct topological regimes: the anomalous Floquet (Fig.~\ref{fig:Fig1}c) with ($W^0$, $W^\pi$)=(1,1) and the Haldane regime (Fig.~\ref{fig:Fig1}d) with (1,0), both supporting topological edge modes with the same chirality.

The overlap between the BEC and the edge mode can be tuned by changing the strength of the harmonic confinement of the tweezer allowing us to distinguish the two regimes with high precision~\cite{supplement,braun_real-space_2024}. Here, we make use of the unique property of anomalous Floquet topological systems which support an edge mode at each quasienergy $\epsilon$, making it extremely robust to the initial-state parameters of the wave packet (Fig.~\ref{fig:Fig1}c). In contrast, the observation of edge modes in the Haldane regime (0-gap) sensitively depends on the initial state (Fig.~\ref{fig:Fig1}d) since it is located at non-zero quasimomentum ($K$ and $K'$ points). By adjusting the tweezer depth we find that a shallow confinement $\omega_s$ suppresses the overlap with the edge mode in the Haldane regime, while maintaining a large overlap with the edge mode in the $\pi$-gap in the anomalous regime. Intuitively, this can be understood by looking at the spatial wavefunction of the edge mode in both regimes. A shallow tweezer generates a BEC at zero quasimomentum, which has good overlap with the $\pi$-gap edge mode in the anomalous regime, but essentially no overlap with the $0$-gap edge mode at finite quasimomentum. In order to further enhance the differential signal we choose a large potential step $V_s$ for generating the interface in the anomalous regime, as it increases the edge-mode velocity, while a shallow potential step $V_d$ is beneficial for large edge-mode velocities in the Haldane regime due to the smaller characteristic energy gap (Fig.~\ref{fig:Fig1}c,d). For experiments where we study edge-mode propagation in the Haldane regime, we utilize a larger radial confinement $\omega_d$, which results in good overlap with the $0$-gap edge mode both in the anomalous-Floquet as well as in the Haldane regime.

\begin{figure*}[!ht]
\centering
\includegraphics{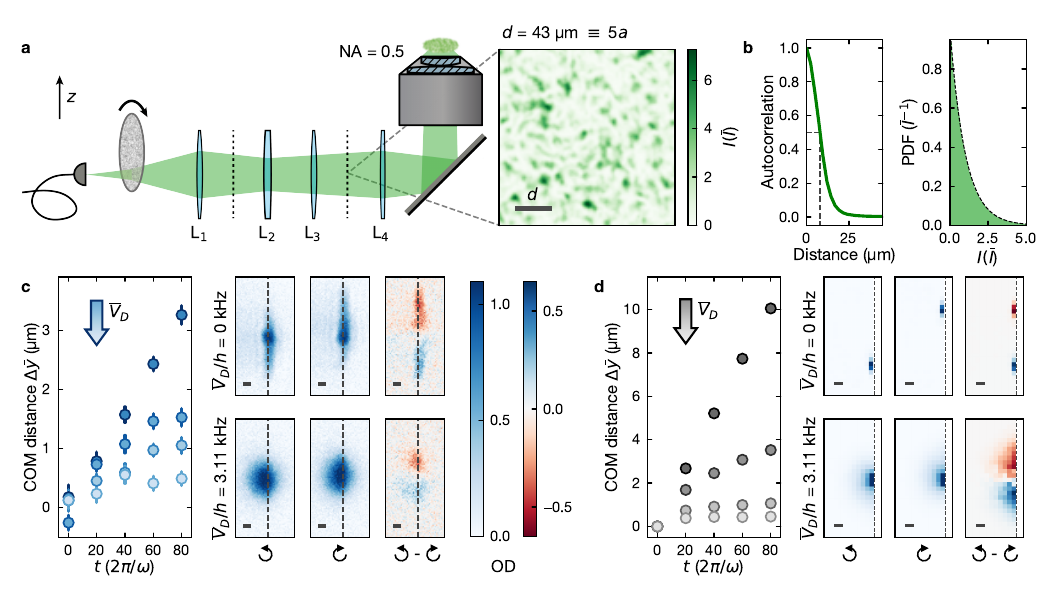}
\caption{\textbf{Optical speckle potential and disorder-induced renormalization of edge-mode propagation in the Haldane regime.}
\textbf{a} Schematic drawing of the optical speckle setup. The uncollimated output of a photonic-crystal-fiber illuminates a small area on a diffuser, scrambling the phase profile of the laser beam. This results in a speckle pattern in the Fourier plane of a lens, which is imaged onto the atoms using two telescopes with a total demagnification of 45. Planes conjugate to the atomic plane are marked by dashed lines. Inset: Measured exemplary optical speckle pattern.
\textbf{b} Left: Radially averaged autocorrelation function with half width half maximum $\SI{7.9}{\micro\meter}$ (dashed lines) evaluated from speckle patterns, as shown in \textbf{a}. We expect the correlation length in the atomic plane to be on the order of $\sigma_r = \SI{0.3}{\micro\meter}$. Right: Expected probability density function (PDF) of a speckle pattern (dashed line) together with a histogram of the speckle intensity evaluated from 118 individual images ($\SI{720}{\micro\meter}\times\SI{720}{\micro\meter}$) before demagnification.
\textbf{c} Experimental data (deep tweezer parameters: $\omega_d$, $V_d$) and \textbf{d} numerical simulation for the edge state propagation in the Haldane regime for $\overline{V}_D / h = $[\SI{0}{\kilo\hertz}, \SI{1.55}{\kilo\hertz}, \SI{3.11}{\kilo\hertz}, \SI{4.67}{\kilo\hertz}] and $\omega/(2\pi) =  \SI{16}{\kilo\hertz}$. The disorder strength is encoded in the color of the data points, going from dark for low disorder strength to light for large disorder strength, as indicated by the arrow shading. For the experimental results, each data point is averaged over 21-23 disorder realizations, for the numerical data we average over 100 disorder realizations. Error bars are extracted via bootstrapping. Right: In-situ pictures after $80T$ for both chiralities (indicated by the arrows) as well as their difference for two disorder strengths $\overline{V}_D / h =[\SI{0}{\kilo\hertz}, \SI{3.11}{\kilo\hertz}]$. The vertical dashed line indicates the location of the topological interface. All scale bars have length 5$a$. In the panels displaying numerical results, the site population is binned into rectangular regions of \SI{1}{\micro\meter} $\times$ \SI{1}{\micro\meter}. The color scale is referenced to the maximum value (half the maximum value) for individual chiralities at $\overline{V}_D / h =\SI{0}{\kilo\hertz}$ ($\overline{V}_D / h =\SI{3.11}{\kilo\hertz}$) and to the maximum absolute value for the difference between chiralities.
} 
\label{fig:Fig2}
\end{figure*}

The experimental sequence starts by generating a BEC of $^{39}$K atoms in a crossed optical dipole trap. By adding a tightly-focused optical tweezer trap at $\lambda_T=\SI{1064}{\nano\meter}$ a tightly-localized BEC of about 200 atoms is formed, whose position is controlled using a 2D acousto-optical deflector (AOD) that allows us to precisely align the position of the BEC relative to the topological interface. Residual atoms not loaded into the tweezer trap are expelled by lowering the large-volume crossed optical dipole trap. While holding the atoms in the tweezer trap, a programmable repulsive optical potential realizing the topological interface is ramped up. The potential is generated using a digital micromirror device (DMD), which is illuminated by a spatially and spectrally incoherent light source with a wavelength centered around $\lambda_S=\SI{638}{\nano\meter}$~\cite{braun_real-space_2024}. Subsequently, a hexagonal optical lattice, generated by three free-running laser beams ($\lambda_L=\SI{745}{\nano\meter}$), interfering under an angle of \SI{120}{\degree}, is ramped up to a depth of 5.9$E_R$ (Fig.~\ref{fig:Fig1}a), where $E_R/h=h/(2 \lambda_L^2 m_\mathrm{K})=\SI{9.23}{\kilo\hertz}$ is the recoil energy, $h$ Planck's constant and $m_\mathrm{K}$ the mass of $^{39}$K. The periodic modulation of the tunnel couplings is realized by varying the intensities of the three laser beams out of phase according to $I_i(t)=I_0(1-m+m\cos(\omega t+\phi_i))$, with $\omega$ the modulation frequency, $m$ the modulation amplitude, $\phi_i=\kappa\frac{2\pi}{3}i$ the phase for the $i$th lattice beam ($i=\{1,2,3\}$), and $\kappa=\pm1$ the chirality of the modulation. After turning on the lattice, the modulation is linearly ramped to its final value $m=0.25$ within $5T$, where $T$ is one modulation cycle. At the end of the ramp, the atoms are suddenly released from the tweezer trap and we monitor the in-situ time evolution of the cloud after a variable time $t$.
The vertical overlap of the tweezer focus and dipole trap is optimized for each experimental run to avoid excitations perpendicular to the 2D lattice, which, together with the dipole trap, has a combined trap frequency $\omega_z/(2\pi) = \SI{0.33 \pm 0.03}{\kilo\hertz}$, and interaction effects are suppressed by setting the scattering length to $6 a_0$, where $a_0$ is the Bohr radius.

\begin{figure*}[!htbp]
\centering
\includegraphics{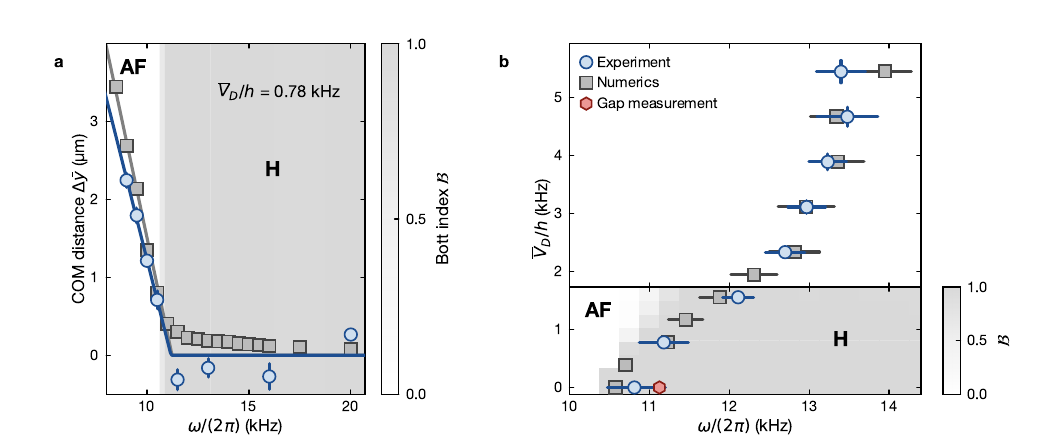}
\caption{\textbf{Measurement of the disorder-driven shift of the topological phase transition.}
\textbf{a} COM distance $\Delta \bar{y}$ as a function of modulation frequency. The solid lines are fits of the function $\max{(-\eta_1\ \omega+\eta_0, 0)}$ ($\eta_1$ and $\eta_0$ are free fit parameters) to extract the phase transition point for the experimental (blue) and numerical data (gray). The latter has been simulated based on a simplified two-band model for 20$T$ and was averaged over 100 disorder realizations. Error bars for experiment and numerics are extracted via bootstrapping. Each experimental data point is the average of at least $n=19$-$20$ measurements per chirality, taken for different disorder realizations and after 50$T$. The gray shading in the background encodes the Bott index $\mathcal{B}$, evaluated for 100 (200) disorder realizations for $\overline{V}_D / h < \SI{1}{\kilo\hertz}$ ($\overline{V}_D / h \geq \SI{1}{\kilo\hertz}$) on $N_s = 576$ lattice sites. The labels (AF) and (H) indicate the anomalous ($\mathcal{B}=0$) and Haldane regime ($\mathcal{B}=1$), respectively. \textbf{b} Location of the topological phase transition obtained from edge state measurements (blue circles), numerical wave packet dynamics in a two-band model (gray squares) and a gap-closing measurement for $\overline{V}_D = 0$ (red hexagon). The horizontal error bars show the error of the fit illustrated in \textbf{a}. The vertical error bar reflects the calibration error of the disorder strength~\cite{supplement}. In the background, the disorder-averaged Bott index is encoded in the color shading up to $\overline{V}_D/h \lesssim \SI{2}{\kilo\hertz}$, where the two bands can still be distinguished numerically.
}
\label{fig:Fig3}
\end{figure*}

To introduce controlled disorder, we generate a repulsive optical speckle potential at a wavelength of \SI{532}{\nano\meter}. Optical speckle potentials are well-established tools in quantum simulation experiments~\cite{billy_direct_2008,Goodman_speckle}, particularly in studies of Anderson localization~\cite{bouyer_quantum_2010,kondov_three-dimensional_2011,jendrzejewski_three-dimensional_2012,semeghini_measurement_2015,lecoutre_bichromatic_2022}, offering disorder with well-characterized statistics and tunable strength over several orders of magnitude. In our setup, a lens transforms a random phase pattern, generated by a holographic diffuser, into a speckle field in the Fourier plane of the lens. This pattern is then projected into the atomic plane using a high-resolution imaging objective with $\mathrm{NA}=0.5$ (Fig.~\ref{fig:Fig2}a). Assuming diffraction-limited performance of the objective, we estimate a speckle correlation length of $\sigma_r = \SI{0.3}{\micro\meter}$ in the horizontal plane and 
$\sigma_z \simeq \SI{2.4}{\micro\meter}$ in the vertical direction, much larger than the harmonic oscillator length in the $z$-direction. 
The intensity distribution of the resulting disorder potential closely follows the expected probability density function $p_{\bar{I}}( I ) =\bar{I}^{-1}\exp[-I/\bar{I}]$,
with $\bar{I}$ being the average intensity~\cite{Goodman_speckle}. Its strength has been calibrated using diffraction measurements~\cite{supplement}, providing the proportionality constant $\alpha_I$ for the potential energy shift seen by the atoms, $\overline{V}_D=\alpha_I \bar{I}$. In the optical setup we further displace the diffuser from the optical axis, such that rotating it results in different potential disorder realizations with similar statistics, as different patches on the diffuser will be illuminated. We use this to average over different disorder realizations.

We start by investigating the propagation of chiral edge modes in the presence of disorder. For weak disorder, it is expected that disorder induces couplings between different quasimomenta within the edge-mode dispersion, resulting in a renormalization of the edge-mode velocity. This is of particular relevance in the context of topological photonics~\cite{guglielmon_broadband_2019,karcher_stability_2024}, where disorder and imperfections can hinder practical applications. 
We focus on the Haldane regime, where the system hosts a single chiral edge mode within the 0-gap. 
The modulation parameters are chosen to lie well within the topological phase, sufficiently far from any phase transition.
To probe the effect of disorder, we measure the difference in the center-of-mass (COM) position $\Delta\bar{y}=\bar{y}_{+1}-\bar{y}_{-1}$ of the wavepacket for both chiralities along the topological interface (aligned along the $y$-direction) at various evolution times, where $\bar{y}_\kappa$ denotes the COM position evaluated from in-situ absorption images, as shown in Fig.~\ref{fig:Fig2}c,d. The in-situ images further reveal that disorder not only slows down edge transport but also induces propagation into the bulk. This is further reflected in a reduced overall transport velocity, evident from the shorter COM distance traveled by the atoms. The velocity observed in the experiment is further suppressed compared to numerical simulations. We attribute this to the finite width of the topological interface and the more finely tuned overlap with the initial state used in the numerics~\cite{stanescu_topological_2010,buchhold_effects_2012,goldman_direct_2013,braun_real-space_2024}.  
We find qualitatively similar behavior in the anomalous regime.

To study the disorder induced shift of the transition between the Haldane and Anomalous Floquet regimes we utilize the preparation scheme introduced in Fig.~\ref{fig:Fig1}c,d.
While the specifics of edge-mode propagation depend on the disorder potential, the presence or absence of chiral motion at the topological interface provides a robust observable for distinguishing between topological regimes. To enhance the contrast between the anomalous Floquet and the Haldane regime, we choose a large potential step $V_s$ in combination with a shallow tweezer $\omega_s$ and measure the COM displacement between the two chiralities after a fixed evolution time, as a function of modulation frequency (Fig.~\ref{fig:Fig3}a).  
In the anomalous regime at lower modulation frequencies, we observe clear chiral motion, which gradually diminishes near the topological phase transition. Beyond a critical frequency, the signal vanishes, marking the transition to the Haldane regime. This behavior is in agreement with numerical simulations. We determine the critical point by fitting a kink function to the experimental and numerical data. 
Note that the absolute value of the COM distances differs between the numerical simulations and the experiment, as described above. 

The robustness of this method for identifying the phase transition is further validated by benchmarking against calculations of the Bott index $\mathcal{B}$~\cite{loring_disordered_2011,Toniolo2022-ie} -- a topological invariant equivalent to the Chern number that is particularly suited to study disordered systems. 
Here, we compute the Bott index for the lowest band of a simplified two-band model, as described in detail in Ref.~\cite{supplement}. 
We identify regions with disorder-averaged Bott index $\mathcal{B}=1$ as the Haldane regime, where $\mathcal{C}^{\mp}=\pm 1$, and regions with $\mathcal{B}=0$ as the anomalous regime, where $\mathcal{C}^{\mp}=0$.
A reliable evaluation of the Bott index requires the two bands to be spectrally distinct, which restricts the accessible disorder strengths to $\overline{V}_D/h \lesssim \SI{2}{\kilo \hertz}$.
In this disorder range, the phase boundary extracted from the Bott index agrees well with the one obtained independently by time-propagating a wavepacket in the same two-band tight-binding model on a finite lattice, evaluated analogously to the experimental data.
The wavepacket-propagation approach remains applicable at larger disorder strengths, where the Bott index cannot be reliably evaluated, and agrees well with the experimental phase boundary over the full disorder range explored. Its agreement with the Bott index at weak disorder further validates the edge-mode observable as a reliable probe of the topological phase transition.
We further benchmark our method for $\overline{V}_D=0$ via comparison to standard gap-closing measurements based on St\"uckelberg interferometry (red data point in Fig.~\ref{fig:Fig3}b)~\cite{zenesini_observation_2010,kling_atomic_2010,Wintersperger2020}.  
This measurement determines the location of the phase transition at \SI{11.13\pm 0.08}{kHz}. The small systematic shift between this gap-closing measurement and the numerical simulations based on a simple two-band model (Bott index in Fig.~\ref{fig:Fig3}b) is due to higher-band mixing present in the real system, which the two-band model neglects.
In the clean limit, we identify the phase transition at \SI{10.92}{\kilo\hertz} by means of a 6-band momentum space model, in good agreement with the experimental results. Extending this model to include disorder is beyond the scope of this work. 

Tracking the phase transition as a function of the applied disorder potential, we observe that the transition shifts to higher modulation frequencies with increasing disorder (Fig.~\ref{fig:Fig3}b), suggesting that disorder favors the anomalous Floquet topological regime, in agreement with theoretical predictions~\cite{hofstetter_phase_transition}. The measured phase boundary is well reproduced by the wavepacket-propagation simulations, including at disorder strengths beyond the range accessible to the Bott index calculation. Numerical studies further suggest that the observed transition shift is quite generic, with only a weak dependence on the correlation length of the disorder potential.

\begin{figure}[!t]
\centering
\includegraphics{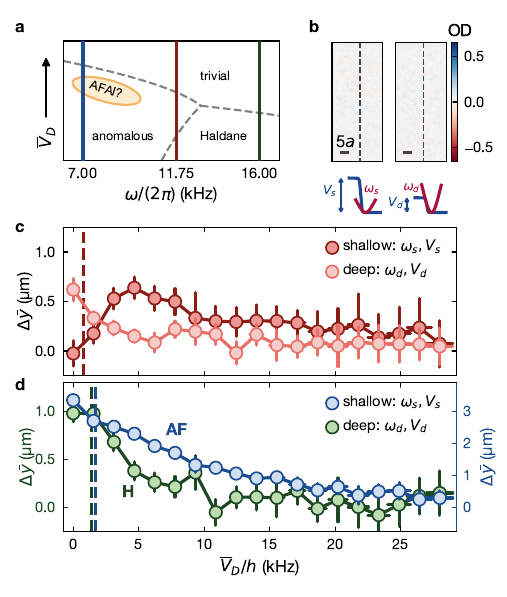}
\caption{\textbf{Topological phase diagram for large potential disorder.} 
\textbf{a} Schematic phase diagram illustrating three different phases: anomalous Floquet, Haldane and topologically trivial. The orange ellipse indicates the potential region of an AFAI phase. The solid lines indicate the parameter regime for the experimental data shown in \textbf{c,d}.
\textbf{b} Difference of absorption pictures obtained for the two different chiralities $\kappa=\pm 1$ (averaged over 100 disorder realizations)
after $50 T$ at $\overline{V}_D / h= \SI{37.32}{\kilo\hertz}$ for the two experimental settings illustrated below.
\textbf{c,d} COM distance after 50$T$ for variable disorder strength and different modulation parameters. The dashed vertical lines indicate the corresponding minimal energy gap without disorder obtained from a 6-band calculation.
\textbf{c} Re-entrance behavior of the edge mode signal for two different experimental settings optimized for the anomalous (dark red) and Haldane (light red) regime, respectively. The modulation frequency [$\omega/(2\pi)=\SI{11.75}{\kilo\hertz}$] is chosen to lie close to the phase transition at zero disorder. 
\textbf{d} Transition to the topologically-trivial regime deep in the anomalous [blue, $\omega/(2\pi) = \SI{7}{\kilo\hertz}$] and Haldane regime [green, $\omega/(2\pi)=\SI{16}{\kilo\hertz}$]. Error bars on $\Delta\bar{y}$ have been extracted by bootstrapping. The error bars in $\overline{V}_D$ give the calibration error of the disorder strength. The blue trace has been plotted on a separate axis on the right due to the much larger COM distance traversed. Each data point is the average of at least $n=19$-$20$ measurements per chirality, with different disorder patterns applied.
}
\label{fig:Fig4}

\end{figure}

Anomalous Floquet systems are predicted to exhibit a rich phase diagram at large disorder strengths (schematic phase diagram: Fig.~\ref{fig:Fig4}a).  
In particular, the interplay between periodic driving and disorder can stabilize genuine out-of-equilibrium topological phases such as Anomalous Floquet Anderson insulators (AFAIs)~\cite{titum_anomalous_2016,kundu_quantized_2020,hofstetter_AFAI}, where a fully localized bulk coexists with an extended topological edge mode. More generally, one expects a transition to the topologically trivial regime once disorder strengths become comparable to the bulk energy gap. Experimentally, this trivial phase is characterized by the disappearance of chiral motion, independent of the initial-state parameters of the BEC (Fig.~\ref{fig:Fig4}b). To probe this phase diagram, we monitor the chiral signal while varying the disorder strength at fixed modulation parameters. Cuts through the phase diagram are illustrated in Fig.~\ref{fig:Fig4}a. Starting near the phase transition in the Haldane regime we find a re-entrant behavior of chiral motion, signaling a transition from the Haldane to the anomalous regime at weak disorder (Fig.~\ref{fig:Fig4}c). We note that a direct transition into a Haldane-like regime with winding numbers $(W^0,W^\pi)=(0,1)$ would instead require both quasienergy gaps to close simultaneously at the same disorder strength and is therefore not generic. Moreover, the persistence of the chiral signal in the deep tweezer configuration, which probes the $0$-gap edge mode (Supplementary Fig.~\ref{fig:1175}), further corroborates the transition into the anomalous regime, in agreement with theoretical predictions~\cite{hofstetter_phase_transition}.
For large disorder, the chiral signal gradually disappears, which we interpret as a transition to the topologically-trivial phase. This dataset was taken with the optimal parameters for the anomalous regime ($V_s$, $\omega_s$). Taking the same scan for a deep tweezer $\omega_d$ and shallow potential step $V_d$, we find a weak signal in the Haldane regime at low disorder strengths, as expected. The signal progressively decays for larger values of $\overline{V}_D$, being further compatible with a transition to a trivial phase in the strong-disorder limit.

In a second set of measurements, we study the behavior deep in the anomalous and Haldane regime where the experimental parameters have been chosen to maximize the signal in each respective regime (Fig.~\ref{fig:Fig4}d). We find that in both cases the chiral signal gradually vanishes and it appears that the transition to the trivial regime happens already at much lower values for the Haldane compared to the anomalous regime despite the fact that the minimal energy gap in the Floquet spectrum is comparable (dashed vertical lines in Fig.~\ref{fig:Fig4}d). Previous work indicates that the relevant energy scale in the anomalous regime is rather determined by the width of the Floquet Brillouin zone~\cite{braun_real-space_2024}, i.e., $\omega/(2\pi)=7\,$kHz, which is much larger than the minimal gap. Note that the vertical scale is different for both data sets for better visibility. This observation further supports theoretical work suggesting an enhanced robustness of anomalous Floquet topological insulators (AFTIs) compared to conventional Chern insulators~\cite{Fleury_Floquet_2016, Peng_Experimental_2016, Rodriguez_Topological_2019, Zhang_Superior_2021}.

In conclusion, edge modes provide a powerful probe for exploring topological phase transitions. While energy-selective excitation of edge and bulk modes is relatively straightforward in photonic systems~\cite{guglielmon_broadband_2019,karcher_stability_2024}, no practical counterpart exists for neutral atoms in optical lattices.
In this work, we demonstrate a novel method for selectively preparing chiral edge modes by tailoring the properties of the initial wavepacket to enhance the overlap with the localized mode at the interface. We have benchmarked the robustness of this approach through numerical simulations and comparisons with the Bott index. We further presented the first experimental investigation of disorder-driven phase transitions between two distinct topological phases in synthetic quantum systems, which are inaccessible using conventional observables ~\cite{cooper_topological_2019}. In the future, these concepts could be extended through energy-resolved preparation schemes, where a cold reservoir is coupled to an initially empty lattice and the injection energy is tuned via the relative chemical potential~\cite{wang_2024}. When combined with microscopic control over initial-state preparation~\cite{gross_quantum_2021,impertro_local_2024}, this opens exciting new avenues for studying the dynamics and robustness of topological edge modes, with particular relevance for applications in topological photonics~\cite{guglielmon_broadband_2019,karcher_stability_2024}.

Exploring topological phases of matter in the presence of disorder further opens new avenues for investigating exotic Floquet topological phases. AFTIs share key features with conventional Chern insulators, such as quantized chiral edge states. However, AFTIs also support novel phases, where all bulk states are localized coexisting with extended edge modes, so-called AFAIs~\cite{titum_anomalous_2016,kundu_quantized_2020,hofstetter_AFAI}. This unique behavior arises from the protection of topological properties by Floquet symmetry, making AFTIs intrinsically more robust than conventional Chern insulators. AFAIs exhibit remarkable resilience to temporal noise~\cite{timms_quantized_2021,zheng_anomalous_2023}, and initial theoretical work suggests this robustness may extend to interacting systems, potentially enabling many-body localized phases with topologically protected edge modes~\cite{nathan_anomalous_2019}. Neutral atoms offer an ideal platform to study these phenomena, by combining the observation of edge modes~\cite{braun_real-space_2024} with bulk expansion in homogeneous lattices~\cite{schneider_fermionic_2012,ronzheimer_expansion_2013} and tunable interactions. Intriguingly, numerical studies suggest that the modulation scheme employed in our current work indeed supports the existence of an AFAI phase, as schematically illustrated in Fig.~\ref{fig:Fig4}a. Adding an additional perturbation in the form of a homogeneous artificial magnetic field may enable experimental studies of a quantized bulk response~\cite{nathan_quantized_2017} and open the door to experimental studies of the topological magnetoelectric effect of 3D topological insulators~\cite{gavensky_quantized_2025}.

\begin{acknowledgments}
\textbf{Acknowledgments:} We thank W.~Hofstetter and his team for fruitful discussions. This work was funded by the Deutsche Forschungsgemeinschaft (DFG, German Research Foundation) via Research Unit FOR 2414 under project number 277974659. The work was further supported under Germany’s Excellence Strategy – EXC-2111–3908148, by the Alfried Krupp von Bohlen und Halbach foundation and under Horizon Europe programme HORIZON-CL4-2022-QUANTUM-02-SGA via the project 101113690 (PASQuanS2.1).

\textbf{Author Contributions Statement:} A.H. performed the experiments and analyzed the data with contributions from J.A. and M.H. J.A. carried out the numerical expansion simulations and the Bott index calculations. M.H. performed the simulation and evaluation for the disorder calibration. C.B. and M.A. supervised and guided the experimental work. All authors contributed to the interpretation of the results and to the writing of the manuscript.

\textbf{Competing Interests Statement:}
The authors declare no competing interests.

\end{acknowledgments}

\begin{methods}

\textbf{Code availability:}
The code that supports the plots within this paper is available from the corresponding author upon reasonable request.

\textbf{Data availability:}
The data that supports the plots within this paper and other findings of this study are
available at https://doi.org/10.17617/3.P2VDLZ.
\end{methods}

\bigskip
\bibliography{bibliography}

\cleardoublepage

\section*{Supplemental Material}
\setcounter{figure}{0}
\setcounter{equation}{0}
\setcounter{section}{0}
\setcounter{table}{0}

\renewcommand{\thefigure}{S\arabic{figure}}
\renewcommand{\theHfigure}{S\arabic{figure}} 
\renewcommand{\theequation}{S.\arabic{equation}}
\renewcommand{\thesection}{S.\Roman{section}}
\renewcommand{\thetable}{S\arabic{table}}



\begin{itemize}
\item[\ref{sec:supp-sequence}] \hyperref[sec:supp-sequence]{Experimental sequence} \hfill \pageref{sec:supp-sequence}
\item[\ref{supp:disorder_pot}] \hyperref[supp:disorder_pot]{Disorder potential} \hfill \pageref{supp:disorder_pot}
\item[\ref{sec:disorder-cal}] \hyperref[sec:disorder-cal]{Disorder calibration} \hfill \pageref{sec:disorder-cal}
\item[\ref{sec:tweezer-alignment}] \hyperref[sec:tweezer-alignment]{Tweezer alignment routine} \hfill \pageref{sec:tweezer-alignment}
\item[\ref{sec:zero-disorder}] \hyperref[sec:zero-disorder]{Measuring the phase transition for zero} \hfill \pageref{sec:zero-disorder}\\\hyperref[sec:zero-disorder]{disorder}
\item[\ref{sec:data-eval}] \hyperref[sec:data-eval]{Data evaluation} \hfill \pageref{sec:data-eval}
\item[\ref{sec:bott-index}] \hyperref[sec:bott-index]{Bott index evaluation} \hfill \pageref{sec:bott-index}
\item[\ref{sec:edge-propagation}] \hyperref[sec:edge-propagation]{Numerical edge state propagation} \hfill \pageref{sec:edge-propagation}
\item[\ref{sec:1175}] \hyperref[sec:1175]{Edge state signal near the phase transition} \hfill \pageref{sec:1175}
\end{itemize}

\FloatBarrier

\section{Experimental sequence}

\label{sec:supp-sequence}

\begin{figure}[!htbp]
\centering
\includegraphics[width=.5\textwidth]{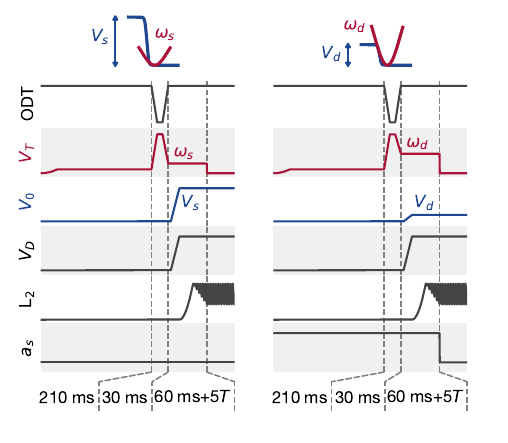}
\caption{\textbf{Experimental sequence.} Illustration of the experimental sequence. The differences for the two different initial-state-preparations are illustrated (left: shallow, right: deep). ODT denotes the intensity of the crossed optical dipole trap, $V_T$ the intensity of the optical tweezer beam, $V_0$ the intensity of the hard-wall potential, $\overline{V}_D$ the intensity of the disorder potential, $L_2$ is the intensity of one of the three lattice beams, and $a$ is the scattering length.}
\label{fig:sequence}
\end{figure}

To optimize the signal when probing the edge states in both the Haldane as well as in the anomalous regime, it is crucial to optimize the number of atoms transferred into the edge state, as well as the edge state velocity. The differences in the preparation protocol for the edge mode in the anomalous (left) and in the Haldane (right) regime are illustrated in Fig.~\ref{fig:sequence}. In both cases, we load the atoms from a BEC of $\mathrm{^{39}K}$ into the optical tweezer within $210\,\si{\milli \second}$. Subsequently, within $30\,\si{\milli \second}$ the tweezer power is briefly increased, while the dipole trap power is reduced, such that atoms not trapped in the tweezer are expelled. Afterwards, the tweezer power is ramped down to its final value, i.e., to a radial confinement frequency $\omega_s/(2\pi) = \SI{1.3\pm0.1}{\kilo\hertz}$ or $\omega_d/(2\pi)=\SI{2\pm0.1}{\kilo\hertz}$ depending on the measurement, as discussed in the main text. From now on the tweezer potential is kept constant, while the other potentials are being ramped up.

First, we ramp up both the potential step used to generate the topological interface $V_0$ generated by a DMD simultaneously with the disorder potential $\overline{V}_D$ within $30\,\si{\milli \second}$. As the edge state velocity  depends on the height of the topological interface, we ramp it to $V_d/h=\SI{2.8}{\kilo\hertz}$ for the Haldane regime, and to $V_s/h=\SI{13.2}{\kilo\hertz}$ for the anomalous regime, to maximize the edge state velocity and hence the signal-to-noise~\cite{braun_real-space_2024}.

Subsequently, we ramp up the optical lattice to a lattice depth of $5.9{E_R}$ within $\SI{30}{\milli\second}$. Finally, we ramp the amplitude of the lattice modulation $m$ to its final value within five modulation cycles. As a last step, we reduce the scattering length one \si{\milli\second} before releasing the atoms from $13a_0$ to $6a_0$ when preparing the edge state in the Haldane regime, while it was constant at $6a_0$ for the preparation scheme in the anomalous regime. Keeping the scattering length at a higher value reduces atom loss due to the stronger confinement at a trap frequency of $2$\,\si{\kilo\hertz}. To initiate the dynamics in the periodically-modulated lattice, we abruptly switch off the tweezer potential.

\section{Disorder potential}
\label{supp:disorder_pot}

\begin{figure}[!htbp]
\centering
\includegraphics{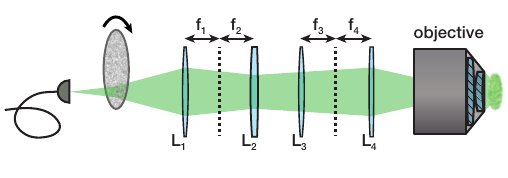}
\caption{\textbf{Setup for generating the speckle potential:} A speckle potential is generated by illuminating a diffuser, and optically Fourier transforming it. Afterwards, it is demagnified by two telescopes, formed by $\mathrm{L_2}$ and $\mathrm{L_3}$ as well as $\mathrm{L_4}$ and the objective. Vertical dashed lines indicate the location of planes conjugate to the atoms.}
\label{fig:Speckle_setup}
\end{figure}

The disorder potential is generated by projecting the Fourier plane of a diffuser onto the atoms, yielding a speckle pattern.
For this, a holographic diffuser (Edmund Optics \#35-693, with a \SI{2}{\degree} diffusing angle) is illuminated at $532\,$\si{\nano\meter} with the uncollimated output of an optical fiber (NKT Photonics aeroGUIDE POWER, seeded by a Lighthouse Sprout-G-15W). The diffuser is mounted in a rotation mount, such that it is illuminated off-center. Rotating the mount allows to illuminate different phase patches on the diffuser, yielding different speckle patterns.

The image on the diffuser is Fourier-transformed by L1 ($f_1 = 100\,\si{\milli\meter}$), and subsequently demagnified by a factor 45 by two telescopes ($f_2=150\,\si{\milli\meter}$, $f_3=100\,\si{\milli\meter}$, and $f_4 = 750\,\si{\milli\meter}$ together with our objective with focal length $f_\mathrm{obj} = 25\,\si{\milli\meter}$).

From images recorded in an intermediary image plane we extract a radial correlation length of \SI{7.9}{\micro\meter}, which after demagnification by a factor 30 is consistent with a numerically simulated correlation length of $296\,\si{\nano\meter}$ in the atomic plane. For the numerical estimate, we initialize the beam in the back focal plane of the objective and propagate it to the entrance pupil using a Fresnel propagation kernel. We then clip the wings of the beam according to the aperture of the objective and Fourier transform it to obtain the field distribution in the atomic plane. The simulation takes two input parameters, the envelope and the correlation length in the back focal plane, which determine the properties of the speckle. The envelope was directly measured in the back focal plane, and is compatible with the correlation length of the images obtained in the intermediary image plane. The correlation length of the beam in the back focal plane is inferred from the envelope on the images taken in the intermediary image plane. We attribute an error of $\pm \SI{2}{\milli\meter}$ to the measurement of the beam waist in the back focal plane, which results in a correlation length $\sigma_r = \SI{296(11:7)}{\nano\meter}$.

To estimate the correlation length of the pattern along the optical axis of the objective, we determine an effective numerical aperture $\mathrm{NA}_\mathrm{eff}$ of our objective that corresponds to the computed correlation length $\sigma_r = \SI{296}{\nano\meter}$ using the standard relation~\cite{Goodman_speckle}

\begin{equation}
    \sigma_r = \frac{\lambda}{4 \cdot \mathrm{NA}_\mathrm{eff}}.
\end{equation}

Using the relation between the radial and the axial correlation length via the effective numerical aperture 

\begin{equation}
    \sigma_z \approx 0.89 \frac{\lambda}{\mathrm{NA}_\mathrm{eff}^2},
\end{equation}

we obtain $\sigma_z = \SI{2.4}{\micro\meter}$.

\section{Disorder calibration}
\label{sec:disorder-cal}

\begin{figure}[!htbp]
\centering
\includegraphics{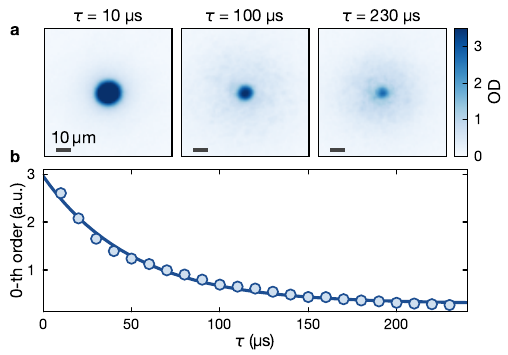}
\caption{\textbf{Speckle diffraction of the BEC. a} Absorption images of the atoms after interaction with a speckle potential with strength $\overline{V}_D/h = \SI{9.3}{\kilo\hertz}$ for different pulse durations $\tau$. The images are taken after 8 ms of expansion in the dipole trap, and averaged over 21 disorder realizations. The scale bar in the bottom left corner indicates \SI{10}{\micro\meter}. \textbf{b} Zeroth order contribution to the atomic density as a function of the speckle pulse duration $\tau$. Each data point is the average over 21 disorder realizations, error bars have been estimated via bootstrapping and are smaller than the marker size. The timescale over which the zeroth order decays is determined through an exponential fit with offset (solid line).}
\label{fig:SpeckleDiffraction}
\end{figure}

To determine the strength of the disorder in our experiment, we devised a calibration scheme for the speckle potential that directly measures the average potential strength $\overline{V}_D$ in the atomic plane, which is crucial for a quantitative analysis of the disorder-driven shift of the topological phase transition. We start out with a BEC consisting of $\approx \num{2e5}$ atoms of $\mathrm{^{39}K}$ at scattering length $6a_0$ in a crossed optical dipole trap with a horizontal trap frequency of $\omega_r / (2\pi) = \SI{25}{\hertz}$, and a vertical trap frequency of $\omega_z/(2\pi) = \SI{0.33(3)}{\kilo\hertz}$. In analogy to Kapitza-Dirac diffraction of a BEC in a pulsed optical lattice \cite{Huckans_KapitzaDirac_2009, Gadway_KapitzaDirac_2009}, we abruptly switch on the speckle potential for a short interaction time $\tau$ between 10\,\si{\micro\second} and 230\,\si{\micro\second}. Subsequently, the atoms are held in the crossed dipole trap for a time of \SI{8}{\milli\second}. One can now observe an exponential decay in atom number from the unscattered atoms as a function of pulse length, which we quantify by performing a two-component fit as described below. By comparison of the measured decay rates to the results of a numerical solution of the Gross-Pitaevskii equation (GPE) the average potential depth of the speckle is inferred.

During the pulse, the atoms are coupled to different momenta according to the Fourier transform $\widetilde V$ of the disorder potential
\begin{equation}
    \langle \vec k |V| \vec{k'}\rangle = \int d\vec x\, V(\vec x)e^{-i(\vec k-\vec{k'})\vec x} = \widetilde V (\Delta \vec k).
\end{equation}
For short pulses ($\tau\overline{V}_D \sim 1$), the final wavefunction in momentum-space is bimodal, which is apparent in the absorption images that are shown in the top panel of Fig.~\ref{fig:SpeckleDiffraction}. The strong, central peak can be attributed to a remnant, unscattered fraction of atoms, which we refer to as zeroth-order contribution. It is surrounded by a broad distribution of scattered atoms that originates from the interaction with the speckle potential. To extract the atom number in the zeroth-order from the experimental images, we proceed as follows. First, we correct for shot-to-shot fluctuations of the atoms in the ODT by centering each image on its peak, which is located as the center of mass of all pixels exceeding a threshold value of 85\% of the maximum OD. Next, the column density is fit to a distribution consisting of a broad Gaussian,
\begin{equation}
    G(x,y) = Ae^{-\frac{1}{2}(x^2/\sigma^2_{x} + y^2/\sigma^2_{y})},
\end{equation}
with fit parameters $A$, $\sigma_x$, $\sigma_y$ for the scattered fraction, and a narrow Thomas-Fermi (TF) profile,
\begin{equation}
    \mathrm{TF}(x,y) = B\left[1-(x/R_x)^2-(y/R_y)^2\right]^{3/2},
\end{equation}
with fit parameters $B$, $R_x$, $R_y$ for the unscattered atoms, set to 0 outside the Thomas-Fermi radius $R_{x,y}$. The atom number up to a missing proportionality constant is then obtained by summing the TF component.

Due to significant overlap between the central component and the wings at weak disorder strength, we are using an expectation-maximization-like scheme to refine the fits. As starting point, a Gaussian is fitted to the wings of the distribution, excluding a central disk of fixed radius. Subsequently, the Thomas-Fermi profile is fit to the residual after subtracting the Gaussian fit, yielding initial estimates of the zero-order amplitude and radii. From then on, the two components are iteratively refined as follows. In each iteration, the pixel-wise responsibilities $w$ are computed as the fractional contribution of each component to the current model
\begin{align}
    w_{\mathrm{TF}}(x,y) &= \frac{\mathrm{TF}(x,y)}{G(x,y)+\mathrm{TF}(x,y)},\\
    w_\mathrm{G}(x,y) &= 1-w_{\mathrm{TF}}(x,y).
\end{align}
Using the responsibility-weighted inverse variance as fit weights, first the Gaussian, then the TF profile are re-fit to the data minus the other component. A shared offset is updated at each iteration as the mean residual over all pixels. The fitting routine terminates once the parameters show no significant change between iterations.

As an additional constraint, the upper bound on the Thomas-Fermi radius at fixed disorder strength is set by the fitted radius of the preceding pulse duration. This constraint enforces the physical expectation that the Thomas-Fermi radius can only shrink with increasing pulse duration. As atoms are progressively scattered out of the central BEC, the remaining atoms adapt to the reduced density in the center of the trap. 

The lower panel of Fig.~\ref{fig:SpeckleDiffraction} shows the zeroth-order contribution for different pulse lengths. It decays exponentially with the duration of the pulse, with a decay rate $\Gamma$ that depends on the average potential strength. To determine the scaling of the decay rate with the strength of the disorder, we use a split-step Fourier algorithm to numerically integrate an effective two-dimensional GPE, suitable for traps with large aspect ratio \cite{Effective2DGPE}.
The simulation is structured to mimic the experimental sequence. First, the ground state in the trap is obtained by means of imaginary-time propagation \cite{ImaginaryTimeAlgorithm}. Afterwards, it evolves in real time in the combined potential of ODT and speckle for a variable pulse duration, before a final $T$/4-expansion in the dipole potential, mapping momenta to positions, separates the zeroth-order from the scattered atoms. The unscattered fraction after time-evolution is extracted with the same fitting routine as for the experimental data, applied to the modulus square of the final wavefunction.

The simulations use a grid of $1512\times1512$ points for $\overline{V}_D/h<$ \SI{4.5}{\kilo\hertz} and $4096\times 4096$ for $\overline{V}_D/h\geq$ \SI{4.5}{\kilo\hertz}. The grid size has to be increased because stronger disorder couples the condensate to higher momenta, which map to larger positions during the T/4 expansion. In both cases, the grid spacing is fixed at $dx=$ \SI{0.061}{\micro\meter}, resolving the disorder correlation length. The radial trap frequency $\omega_r$ enters directly into the simulation via the harmonic confinement, while atom number, scattering length and axial trap frequency $\omega_z$ enter parametrically through a density dependent local chemical potential. For the calibration curve, we evaluate the decay of the zeroth-order for 15 disorder strengths spaced logarithmically from \SI{0.5}{\kilo\hertz} to \SI{50}{\kilo\hertz}, averaging 25 disorder realizations per value. The final continuous curve is obtained by interpolating the numerical decay rates using a piecewise cubic Hermite interpolating polynomial.

For the parameter regime in which our experiment operates, interactions have a negligible effect on the calibration. We verify this by comparing to a simulation of noninteracting particles, which yields a calibration curve that agrees with the GPE result over the relevant range of disorder strengths. This is consistent with the interaction energy per particle extracted from the GPE simulation, approximately \SI{0.5}{\kilo\hertz}, which is well below the smallest disorder strength of \SI{3}{\kilo\hertz} used throughout this work. Only once the two become comparable do we see a clear impact of the interactions, with a larger decay rate in the interacting case compared to the single-particle case.

\begin{figure}[!t]
\centering
\includegraphics{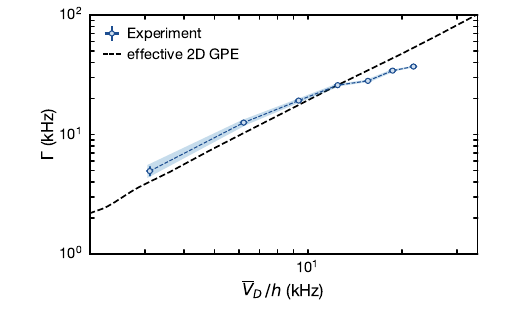}
\caption{\textbf{Experimental disorder calibration.}
The proportionality constant $\alpha$ connecting photodiode voltage to disorder strength in the atomic plane is obtained by fitting the experimentally measured decay rates (blue dots) to the numerically obtained rates (dashed black line). The shaded region around the data marks the systematic uncertainty from the OD thresholding during evaluation as described in the text. The errorbars show the statistical error obtained from bootstrapping, the $x$-error originates from the calibration constant and is not independent per point. Each experimental data point is the fitted decay constant from a measurement as illustrated in Fig.~\ref{fig:SpeckleDiffraction}b, consisting of 23 data points, each being the average over 21 different disorder realizations.}
\label{fig:DisorderCal}
\end{figure}

In the experiment, the power of the speckle beam is stabilized, the disorder strength $\overline{V}_D$ in the atomic plane is directly proportional to the set value $U_\mathrm{PD}$: $\overline{V}_D /h = \alpha \cdot U_\mathrm{PD}$. To find the proportionality constant $\alpha$, we perform a single parameter fit of the experimental data to the numerically calculated calibration curve (Fig.~\ref{fig:DisorderCal}). This procedure corresponds to a rescaling of the horizontal axis of the plot, while leaving the vertical axis with the measured decay rates unchanged. For the specific case of our optical setup, we obtain $\alpha = \SI{31.1}{\kilo\hertz\per\volt}$.

Due to the high density of the BEC in the center of the ODT, a reliable measure of the atomic density is difficult with standard absorption imaging techniques and we see a clear deviation between measured optical depth (OD) and expected TF profile of the condensate once the OD becomes too large. To account for this error, we introduce an OD threshold to the experimental fitting routine, excluding the high-density region from the fit. Varying this threshold over the range OD=3.25 to 4.5, which spans the region where the measured shape starts to deviate from a TF profile up to where no points are excluded, we find that the calibration constant is robust against its precise value. We estimate a systematic error $\Delta\alpha_\mathrm{sys}=\pm0.5\,\si{\kilo\hertz\per\volt}$ on the calibration constant from the spread of values throughout the scan, taking the minimum and maximum as bounds and stating the central value.
The statistical error is obtained by bootstrapping the zero-order fit. For each of $N=500$ realizations we resample the images at every hold time with replacement, recompute the zero-order atom number, and fit the decay constant $\Gamma$. Fitting the calibration constant to these values yields one $\alpha$ per realization and we take the standard deviation over all realizations as the statistical error $\Delta\alpha_\mathrm{stat}=\pm0.6\,\si{\kilo\hertz\per\volt}$. We obtain for the final calibration constant $\alpha = \left(31.1 \pm \left(0.6\right)_\mathrm{stat} \pm \left(0.5\right)_\mathrm{sys} \right)\,\si{\kilo\hertz\per\volt}$.

\section{Tweezer alignment routine}
\label{sec:tweezer-alignment}

\begin{figure}[!htbp]
\centering
\includegraphics{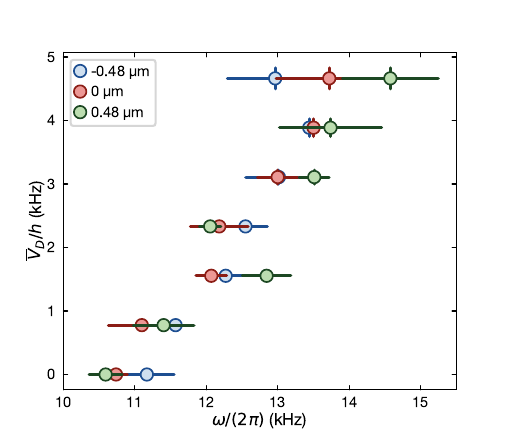}
\caption{\textbf{Tracking the phase transition at different positions.} Location of the phase transition for varying disorder, data taken at three different locations perpendicular to the wall. Each data point is the average over 5 to 7 disorder realizations. The error bars are the fit error, analogous to Fig.~\ref{fig:Fig3} in the main text.}
\label{fig:3pos}
\end{figure}

Before each experimental run, the position of the tweezer was aligned to the potential step by varying the tweezer position with respect to the step, optimizing the strength of the edge state signal. As different wall heights are required to obtain a large signal in the Haldane- and anomalous regime, the tweezer position was optimized separately to deliver a strong signal at $\omega/(2\pi)=7$\,\si{\kilo\hertz} for the shallow tweezer and at $\omega/(2\pi)=16$\,\si{\kilo\hertz} for the deep tweezer, as at these modulation frequencies the energy gaps are reasonably large. In the following, data was taken at the location, where the absolute value of the difference signal $\abs{\Delta\mathrm{OD}}$ integrated in a region close to the edge of the system was maximized. Here, for data taken with the shallow tweezer settings $(\omega_s, V_s)$ the position was chosen where the edge state signal at $\omega/(2\pi)=\SI{7}{\kilo\hertz}$ was maximized, and respectively for the deep tweezer settings $(\omega_d, V_d)$ the position where the edge state signal at $\omega/(2\pi) = \SI{16}{\kilo\hertz}$ was maximized.

Furthermore, data was taken for varying tweezer positions to ensure that this alignment procedure does not introduce systematic bias into the data taken. An exemplary measurement is shown in (Fig.~\ref{fig:3pos}), illustrating good agreement of the observed shift of the phase transition between the anomalous and Haldane regime for different positions of the tweezer perpendicular to the wall.

\section{Measuring the phase transition for zero disorder}
\label{sec:zero-disorder}

\begin{figure}[!htbp]
\centering
\includegraphics{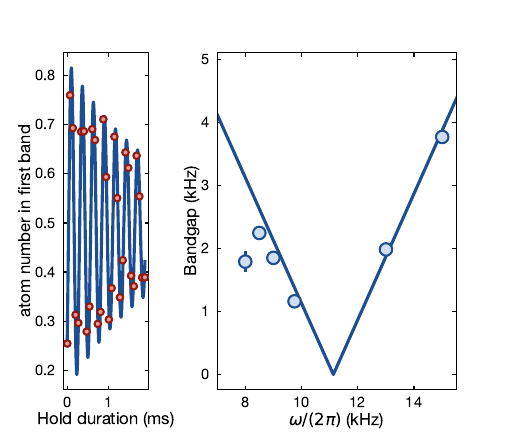}
\caption{\textbf{Measuring the phase transition at $\overline{V}_D = 0$: Left:} Bandgap measured via Stückelberg interferometry at $\omega/(2\pi)=15\,\si{\kilo\hertz}$. Each data point corresponds to one or two measurements. The solid line is a fit of the form $a\cdot e^{-b\cdot t} \cos({\omega t -\phi})$, with $a$, $b$, $\omega$ and $\phi$ being fit parameters, and $\omega/(2\pi)$ being the bandgap. \textbf{Right:} After tracking the energy gap at different modulation frequencies, we perform a fit of the form $\abs{\omega-\omega_{c}}$, and obtain $\omega_c/(2\pi)=(11.13\pm0.08)\,\si{\kilo\hertz}$. Error bars are the fit error for the bandgap fit.}
\label{fig:gap}
\end{figure}

To benchmark the numerical simulations of the Bott index as well as the location of the phase transition extracted from the edge state evolution, we compare it to a gap-closing-measurement at $\overline{V}_D=0$\,\si{\kilo\hertz}. For this, we employ Stückelberg interferometry~\cite{zenesini_observation_2010,kling_atomic_2010,Wintersperger2020}:

We load a bulk BEC at scattering length $6a_0$ into our optical lattice, and, while ramping up the amplitude modulation, nonadiabatically sweep the phase of two lattice beams to accelerate the atoms to the next $\Gamma$ point in quasimomentum space. This nonadiabatic sweep leads to a splitting of the atom population onto the lowest two bands of our system. Subsequently, we hold the atoms at this quasimomentum for an integer number of modulation periods, and afterwards ramp lattice phase and amplitude modulation back to transfer the atoms to zero quasimomentum (at the $\Gamma$-point). Ramping the phase back nonadiabatically recombines the population in the two bands, leading to oscillations in the band population due to the acquired phase difference, as illustrated on the left in Fig.~\ref{fig:gap}.

This means, that in the vicinity of a phase transition at $\Gamma$, we can observe the size of the gap reduce to zero, and increase again (right side in Fig.~\ref{fig:gap}). From this we extract the phase transition for vanishing disorder $\overline{V}_D=0\,$\si{\kilo\hertz} at $\omega_c/(2\pi)=\SI{11.13\pm0.08}{\kilo\hertz}$, where the error bar denotes the fit error.  From a 6-band model calculation, we determine a transition frequency of \SI{10.92}{\kilo\hertz}.

\section{Data evaluation}
\label{sec:data-eval}

To track the movement of the edge states, we calculate the center-of-mass (COM) position of the atoms. To this end, we calculate the weighted average of the position, using the optical density extracted from absorption pictures as a weight:

\begin{equation*}
\overline{y} = \frac{\sum_i y_i w_i}{\sum_i w_i} ,
\end{equation*}

where $y_i$ denotes the position along the topological interface each camera pixel $i$ corresponds to in the atomic plane, and $w_i$ is the optical density (OD) at this pixel. The absorption pictures have been calibrated to account for saturation effects due to the non-negligible intensity of the imaging beam with respect to the saturation intensity of the atoms~\cite{Reinaudi:07}.

To track the chiral movement from this, we evaluate the difference between the position obtained for modulation with chirality $\kappa = 1$ and for modulation with chirality $\kappa = -1$ parallel to the step of the hard-wall potential.

To estimate the error in this determination, we employ bootstrapping: From the set of all absorption pictures taken with the same parameters, we choose random sets of the same length as the initial set, such that images can be selected multiple times. We perform the evaluation for 100 of such sets, and take the standard deviation in the results to be the error.

\section{Bott index evaluation}
\label{sec:bott-index}

For disordered systems, lacking translational invariance, conventional topological invariants defined in terms of geometric properties of Bloch states in momentum space are no longer practical. The Bott index, introduced by Loring and Hastings \cite{loring_disordered_2011}, is a real-space topological marker that can be evaluated on a disordered lattice to quantify whether topological features prevent the existence of a set of localized wavefunctions spanning a band's eigenspace. For vanishing disorder, the Bott index of a given energy band has been shown to coincide with its Chern number in the thermodynamic limit \cite{Toniolo2022-ie}. In the following, we adopt the procedure outlined in Ref.~\cite{loring_disordered_2011} to evaluate the Bott index for the lower band of the effective Hamiltonian resulting from our periodic modulation scheme.

\begin{figure}
\centering
\includegraphics{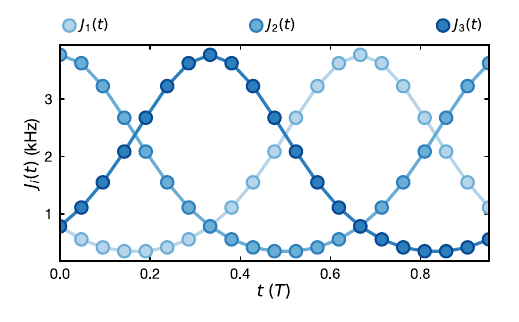}
\caption{\textbf{Temporal discretization of nearest-neighbor tunneling modulation.} The continuous time-dependence of the tunneling rates along the three main directions of the honeycomb lattice is numerically approximated by evaluation at $N=21$ equidistant timesteps across the modulation period.} 
\label{fig:modulation}
\end{figure}

\begin{figure}
\centering
\includegraphics{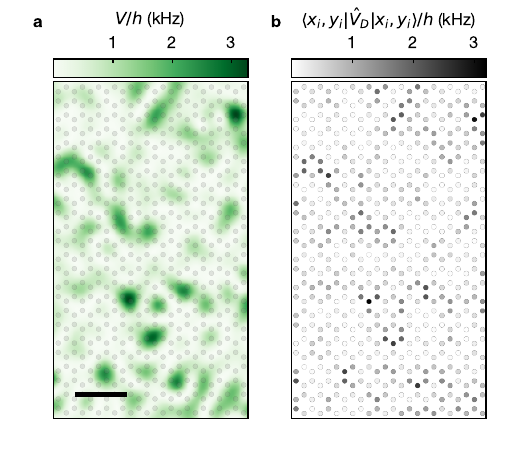}
\caption{\textbf{Spatial sampling of on-site energy offsets from a speckle pattern.} \textbf{a}, Potential landscape from a numerical speckle pattern with $\overline{V}_D/h=0.5$\,\si{\kilo\hertz} and $\sigma_r = \SI{296}{\nano\meter}$ over a region covered by a finite lattice consisting of $N_s=576$ sites. The scale bar covers a distance of $5a$. \textbf{b}, Corresponding on-site energy offsets sampled by the lattice sites, contributing the same diagonal terms to the instantaneous tight-binding Hamiltonian at every timestep, as for an effectively static disorder potential.}
\label{fig:sampling}
\end{figure}

The sinusoidal drive of the intensity of the three laser beams interfering to form the optical honeycomb lattice results in the following time dependence for the nearest-neighbor tunneling amplitudes:
\begin{equation}
    J_i(t) = Ae^{B\cos(\omega t + \phi_i)}+C,
\end{equation}
where $\phi_i=\frac{2\pi}{3}(i-1), i \in \{1,2,3\}$ and $A$, $B$ and $C$ are parameters that depend on the modulation amplitude $m$. For $m=0.25$, as set for all experimental results discussed in the main text, $A \approx 0.220E_R$, $B \approx 0.767$ and $C \approx -0.065E_R$, with the recoil energy $E_R = h^2/(2 \lambda_L^2 2m_{\textrm{K}}) \approx h \times 9.23\,\si{\kilo\hertz}$. Next-nearest neighbor tunneling amplitudes are about one order of magnitude smaller, and thereby neglected in the following. In order to evaluate the effective Hamiltonian numerically, the modulation period is discretized into $N$ timesteps to evaluate $J_i(t_j)$ and the corresponding instantaneous Hamiltonian $\hat{H}(t_j)$ for $t_j = \frac{T}{N}j, \: j =0,1,...,N-1$. For all numerical results discussed in the main text, the number of timesteps is set to $N=21$, resulting in the discretization of the tunneling rates along the three lattice directions shown in Fig.~\ref{fig:modulation}. At each timestep, $\hat{H}(t_j)$ is perturbed by an additional term $\hat{V}_{D}$, modeling on-site energy offsets $\langle x_i, y_i| \hat{V}_D | x_i, y_i \rangle$ sampled by the hexagonal lattice on a numerical speckle potential, as displayed in Fig.~\ref{fig:sampling}. To ensure correct ordering of eigenstates upon diagonalization, we subtract the mean of all $N_s$ on-site energy offsets from each diagonal entry of $\hat{H}(t_j)$. 
\\
Ultimately, the time-evolution operator over one modulation period reads 
\begin{equation}
    \hat{U}(T) = \prod_{j=0}^{N-1} e^{-\frac{i}{\hbar}\left(\hat{H}(t_{N-j-1}) + \hat{V}_{D} \right)\frac{T}{N}}, 
\end{equation}
from which the effective Hamiltonian is readily obtained:
\begin{equation}
	\hat{H}_{\text{eff}} = \frac{i \hbar}{T} \ln{(\hat{U}(T)} ).
\end{equation}

To evaluate the Bott index for a given disorder realization, we enforce periodic boundary conditions on a rectangular system consisting of $N_s$ lattice sites, construct $\hat{H}_{\text{eff}}$ in real space and diagonalize it to obtain its eigenbasis $\{ |\psi_{n} \rangle ,\: n = 1,...,N_s\}$. In a second step, we construct the projector onto the lower-band eigenspace $\hat{P}=\sum_{1\leq n\leq N_s/2} |\psi_n\rangle \langle \psi_n | $ to evaluate the following band-projected position operators:
\begin{equation}
	\hat{G}_{\alpha} = \hat{P} e^{\frac{2\pi i}{L_{\alpha}} \hat{\alpha}} \hat{P},\:  \alpha=x,y,
\end{equation} 
with $L_\alpha$ denoting the system size in $x$- or $y$-direction.\\
Finally, the Bott index of the lower band is given by
\begin{equation}
\mathcal{B} = \frac{1}{2\pi} \Imag  \left(\Tr (\ln (\hat{\mathcal{G}}_{y}\hat{\mathcal{G}}_{x}\hat{\mathcal{G}}_{y}^{\dagger}\hat{\mathcal{G}}_{x}^{\dagger}))\right),
\end{equation}
where $\hat{\mathcal{G}}_{\alpha} = \hat{G}_{\alpha} + (\hat{\mathbb{1}}-\hat{P})$.

In the absence of disorder ($\overline{V}_D=0$), the transition from $\mathcal{B}=0$, corresponding to the anomalous regime, to $\mathcal{B}=1$, corresponding to the Haldane regime, takes place at a modulation frequency $\omega / (2\pi) \approx 10.44$\,\si{\kilo\hertz} for $m=0.25$, in agreement with the numerical evaluation of the corresponding Chern number in a two-band model with the same modulation scheme. Numerical estimates for transition frequencies provided by a two-band model, which do not capture higher-band corrections, systematically underestimate the transition frequency predicted by a six-band calculation, namely $\omega / (2\pi) \approx 10.92$ \si{\kilo\hertz}, and the one measured experimentally, as shown in Fig.~\ref{fig:Fig3} in the main text and studied in detail in Ref.~\cite{Wintersperger2020}.

\begin{figure}
\centering
\includegraphics{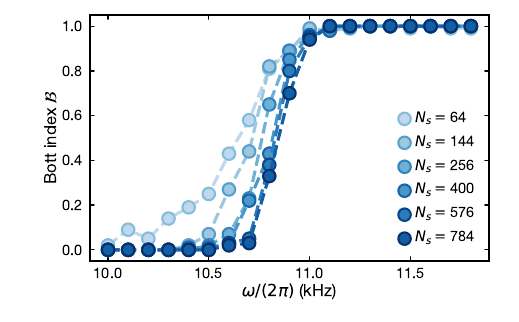}
\caption{\textbf{Finite-size effects on average Bott index evaluation.} For increasing number of lattice size $N_s$, the frequency at which the transition between the anomalous ($\mathcal{B}=0$) and the Haldane regime ($\mathcal{B}=1$) occurs for a disorder strength of $\overline{V}_D / h = 1$\,\si{\kilo\hertz} converges to a fixed value for system sizes larger than $N_s=400$. The Bott index is averaged over $N_\text{avg}=100$ disorder realizations for every combination of modulation frequency and system size.}
\label{fig:convergence}
\end{figure}

The number of lattice sites set for the Bott index evaluations discussed in the main text is $N_s = 576$. Figure \ref{fig:convergence} shows the Bott index across the transition between the anomalous and the Haldane regime for varying system size $N_s$ and fixed disorder strength $\overline{V}_D / h = 1$\,\si{\kilo\hertz}, averaged over $N_{\text{avg}}=100$ disorder realizations. For system sizes larger than $N_s=400$, the average Bott index converges to the thermodynamic limit, which justifies the value chosen for $N_s$ and rules out finite-size effects.

\section{Numerical edge state propagation}
\label{sec:edge-propagation}

The procedure to obtain $\hat{U}(T)$ outlined in the previous section can be employed on a lattice with closed boundary conditions to investigate the propagation dynamics of wavepackets localized to the system's numerical edges for increasing disorder strength. Repeated action of $\hat{U}(T)$ on an initial state $\ket{\varphi_0}$ provides access to the same observables probed experimentally, i.e., the COM distance. After $n$ modulation periods, the initial state is mapped onto
\begin{equation}
    \ket{\varphi(t=nT)} = \hat{U}(nT)\ket{\varphi_0} = \left(\hat{U}(T)\right)^n \ket{\varphi_0}.
\end{equation}

\begin{figure*}[!htbp]
\centering
\includegraphics{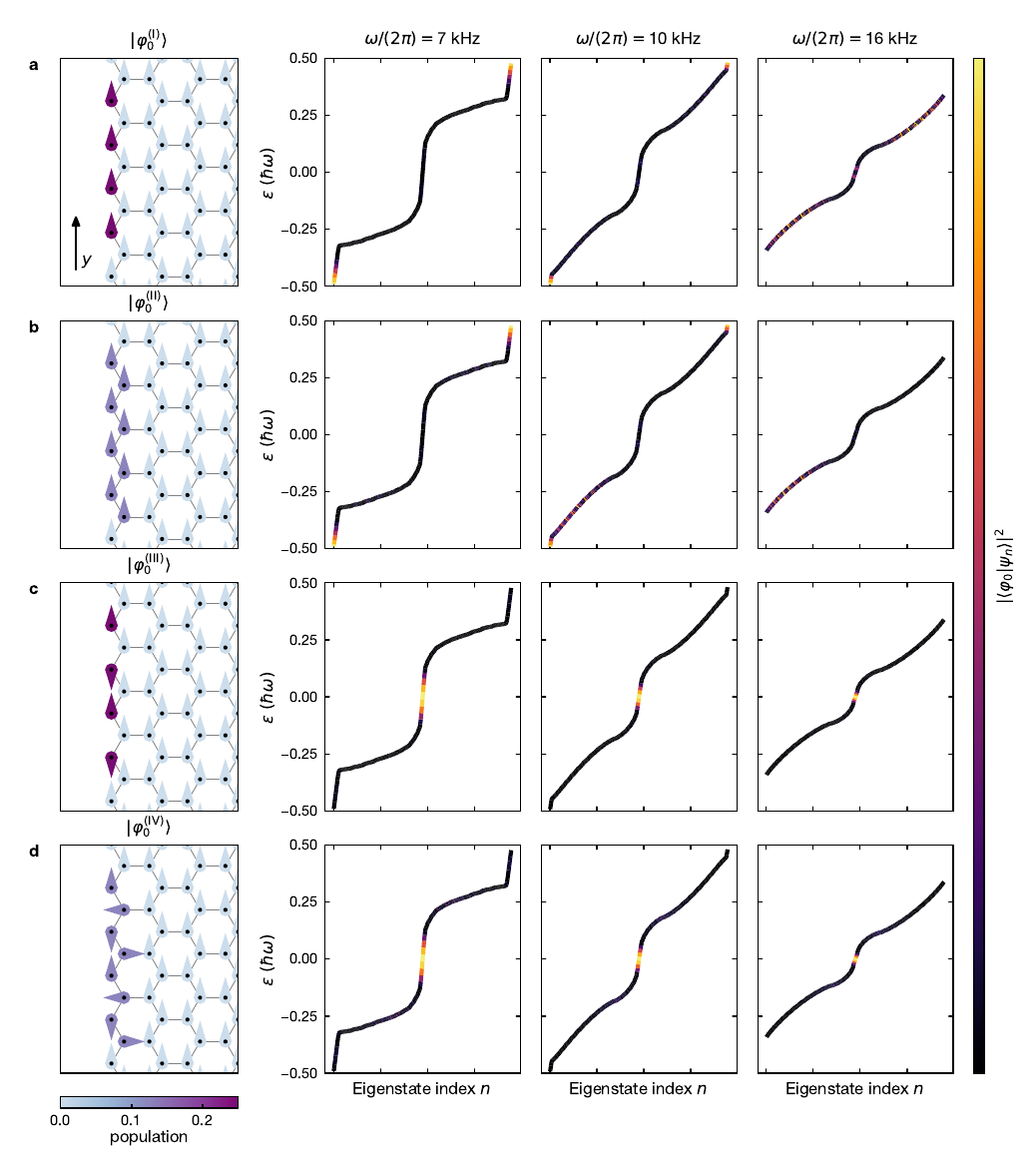}
\caption{\textbf{Numerical edge mode preparation.} Four possible choices for the initial wavefunction and corresponding overlap with the eigenstates $\{ |\psi_{n} \rangle ,\: n = 1,...,N_s\}$ of the effective Hamiltonian on a finite lattice strip deep in the anomalous regime [$\omega/(2 \pi) = 7$\,\si{\kilo\hertz}], close to the transition into the Haldane regime [$\omega/(2 \pi) = 10$\,\si{\kilo\hertz}] and deep in the Haldane regime [$\omega/(2 \pi) = 16$\,\si{\kilo\hertz}]. The little arrows indicate the site-dependent phase of the wave function. \textbf{a}, Equal distribution of the population over four next-nearest neighbors along the edge, belonging to the same sublattice, results in good overlap with the edge state in the $0$-gap of the anomalous regime, but also nonvanishing overlap with the edge state in the $\pi$-gap of the Haldane regime. \textbf{b}, Populating sites on the second sublattice as well significantly reduces the overlap with the edge states in the Haldane regime. \textbf{c, d} The two corresponding options for $\ket{\varphi_0}$ in the Haldane regime result in similar overlaps with eigenstates at each modulation frequency considered. For the sake of visibility, the color limits of each plot displaying overlaps with eigenstates are renormalized to the corresponding maximum value.}
\label{fig:wavefunctions}
\end{figure*}

Depending on the overlap $|\langle \varphi_0 | \psi_n \rangle|^2$ of the initial state with the eigenstates $\{ |\psi_{n} \rangle ,\: n = 1,...,N_s\}$ of the effective Hamiltonian in a given regime, the wavefunction population will propagate along the numerical boundary of the system or scatter towards its bulk. Both in the anomalous and in the Haldane regime, for a zigzag edge geometry, only the outermost lattice sites have a non-zero population for an infinitely high and steep boundary. However, as described in Ref.~\cite{braun_real-space_2024}, the edge state wave functions display a uniform phase pattern in the anomalous regime and an alternating phase pattern in the Haldane regime, such that $\ket{\varphi_0}$ can be initialized to have a good overlap with the edge states in the anomalous regime and vanishing overlap with the edge state in the Haldane regime, providing a numerical equivalent to the regime-sensitive state preparation achieved experimentally by varying the tweezer depth.

To closely reproduce the experimental setup, we generate lattice strips for which the system boundary in the $y$-direction, along which we initialize and time-propagate numerical wavepackets, displays a zigzag termination geometry. Suitable system sizes are chosen depending on the largest expansion time of interest such that edge modes do not propagate past the system's corner. This ensures that edge dynamics can be solely investigated by means of the center-of-mass in the $y$-direction and the time-evolution thereof. For all numerical results displayed in Fig.~\ref{fig:Fig2}d in the main text, probing edge propagation deep in the Haldane regime up to expansion times of $80T$, numerical strips consist of $N_s=3584$ lattice sites, with a ratio between the lengths of the system in the $y$- and $x$-direction of $L_y/L_x \approx 2.7$. On the other hand, for Fig.~\ref{fig:Fig3} of the main text, the fixed expansion time at which the COM distance decay is evaluated across the disorder-shifted transition amounts to $20T$, such that smaller systems with $N_s=952$ lattice sites suffice, with a slightly larger ratio of $L_y/L_x \approx 2.8$. 

Intuitively, to numerically populate edge modes in the $\pi$-gap of the anomalous regime, $\ket{\varphi_0}$ should be prepared to closely match the corresponding wavefunctions, i.e., such that only lattice sites along the boundary, belonging to the same sublattice and spaced by $\sqrt{3}a$, are populated with uniform phase patterns~\cite{braun_real-space_2024}, as displayed in the left panel of Fig.~\ref{fig:wavefunctions}a. The edge mode in the $0$-gap of the Haldane regime is populated with an alternating phase pattern as displayed in Fig.~\ref{fig:wavefunctions}c. For the purposes of the numerics discussed in Fig.~\ref{fig:Fig3} of the main text, however, the wavefunction $|\varphi_0^{(\textrm{I})}\rangle$ in Fig.~\ref{fig:wavefunctions}a is seen to display a nonvanishing overlap with the edge state wavefunctions in the $0$-gap of a system deep in the Haldane regime, $\omega / (2\pi) = 16$\,\si{\kilo\hertz} (rightmost panel of Fig.~\ref{fig:wavefunctions}a), resulting in a residual chiral signal in the Haldane regime. Moreover, we find that the dynamics of such an initial wavepacket is strongly influenced by the initial phase of the modulation. In contrast, we find that $|\varphi_0^{(\textrm{II})}\rangle$, a slightly extended wavefunction that more closely resembles the situation realized experimentally (leftmost panel of Fig.~\ref{fig:wavefunctions}b), provides a more robust observable with higher contrast. Here, also sites slightly away from the interface (on the other sublattice) are populated. This initial state displays good overlap with the edge mode of interest in the anomalous regime and vanishing overlap with the edge mode in the Haldane regime, in qualitative analogy with state preparation with the shallow tweezer. It thereby provides an accurate probe of the disorder-driven shift of the transition and hence a suitable choice for the numerical simulations included in Fig.~\ref{fig:Fig3}.

In the Haldane regime, the transversal width of the initial state $\ket{\varphi_0}$ does not seem to play a significant role. We perform a similar comparison using the initial wavefunctions shown in the leftmost panels of Fig.~\ref{fig:wavefunctions}c,d. Both yield similar results in terms of overlap with eigenstates at $\omega / (2\pi) = 16$\,\si{\kilo\hertz}. In principle, both initial states $|\varphi_0^{(\mathrm{III})}\rangle$ and $|\varphi_0^{(\mathrm{IV})}\rangle$ could be used for the numerical simulations shown in Fig.~\ref{fig:Fig2}d. 
As it more closely resembles the ideal wavefunction of the $0$-gap edge state, we employ $|\varphi_0^{(\mathrm{III})}\rangle$, the first option.

\begin{figure}
\centering
\includegraphics{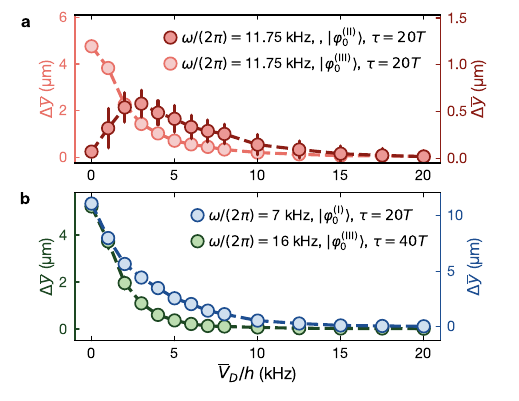}
\caption{\textbf{Numerical edge mode propagation as a function of disorder.} Starting from the zero disorder limit, for modulation frequencies lying deep in the anomalous regime ($\omega/(2\pi)=\SI{7}{\kilo\hertz}$), deep in the Haldane regime ($\omega/(2\pi)=\SI{16}{\kilo\hertz}$) and on the Haldane side of the zero-disorder phase transition ($\omega/(2\pi)=\SI{11.75}{\kilo\hertz}$), we numerically investigate propagation dynamics of edge modes in different energy gaps by means of the same observable considered experimentally, the COM distance $\Delta \overline{y}$. A qualitative correspondence between the initial states employed in the simulations and the experimental tweezer configurations is discussed in Sec. \ref{sec:edge-propagation}. \textbf{a}, COM distance as a function of disorder, starting from the Haldane regime at $\omega/(2\pi)=11.75$~kHz, for two different initial states: $|\varphi^{\textrm{(III)}}_{0}\rangle$ (light red), which overlaps well with the $0$-gap edge mode of the Haldane regime, and $|\varphi^{\textrm{(II)}}_{0}\rangle$ (dark red), which overlaps well with the $\pi$-gap edge mode of the anomalous regime.
\textbf{b}, Disorder-driven breakdown of COM distance starting from the anomalous Floquet [$\omega/(2\pi)=7$ kHz, blue curve] and Haldane regime [$\omega/(2\pi)=16$ kHz, green curve] after an expansion time of 20$T$ and 40$T$, respectively. Every data point in both subplots is obtained by averaging over $N_{\text{avg}}=100$ disorder realizations. Error bars on data points have been obtained by bootstrapping.}
\label{fig:reply_fig}
\end{figure}

\begin{figure}[b]
\centering
\includegraphics{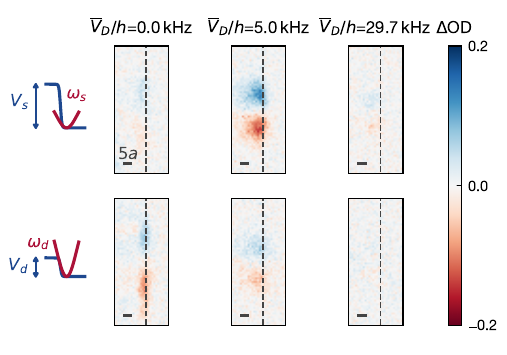}
\caption{\textbf{Differential edge signal near the phase transition.} Real-space images of the differential edge signal at $\omega/(2\pi) = \SI{11.75}{\kilo\hertz}$ for shallow (top) and deep (bottom) tweezer configurations. The data was taken in a separate set of measurements equivalent to Fig.~4c, in order to average over a larger number of disorder realizations at selected disorder strengths $\overline{V}_D = \left[ \SI{0}{\kilo\hertz}, \SI{5.0}{\kilo\hertz}, \SI{29.7}{\kilo\hertz} \right]$. In particular, each image is obtained by taking the difference between the two chiralities and averaging over 146-147 experimental realizations per chirality. From left to right, the disorder strength increases, with $\overline{V}_D/h = \SI{5.0}{\kilo\hertz}$ corresponding to the disorder strength at which the center-of-mass displacement $\Delta\bar{y}$ is maximal in the shallow tweezer configuration. The dashed line indicates the location of the topological interface, and all scale bars correspond to a length of $5a$.}
\label{fig:1175}
\end{figure}

Real-space propagation numerics can provide an insight into the breakdown of topological edge transport in the limit of disorder strengths $\overline{V}_D$ being large enough to induce a Floquet phase transition from a topologically non-trivial regime to a trivial one. As displayed in Fig.~\ref{fig:reply_fig}b, the center-of-mass displacement $\Delta\overline{y}$, evaluated analogously as for experimental absorption images, i.e. by subtraction of the center of mass along the direction parallel to the potential step of the disorder-averaged distributions for the two opposite chiralities, continuously decreases for increasing disorder strength $\overline{V}_D$, regardless of whether in the anomalous regime at $\omega/(2\pi)=7$ kHz or in the Haldane regime at $\omega/(2\pi)=16$ kHz. The steady decay of the chiral signal approaches $\Delta\overline{y} \approx 0$, consistent with a transition to the trivial regime, for significantly weaker disorder in the latter regime compared to the former, underscoring the major robustness of edge modes in the anomalous regime observed experimentally. Similarly as for the experimental data displayed in Fig.~\ref{fig:Fig4}d of the main text, the initial wavefunction $| \varphi_0 \rangle$ propagated in time by $\hat{U}(T)$ is chosen to maximize the overlap with the edge mode of interest in the corresponding regime: four sites on the sublattice to which the outermost sites belong are populated with the uniform and alternating phase patterns, corresponding to the initial states $|\varphi_0^{(\textrm{I})}\rangle $ and $|\varphi_0^{(\textrm{III})}\rangle $ illustrated in the left panels of Fig.~\ref{fig:wavefunctions}a, c at $\omega/(2\pi)=7$ kHz and $\omega/(2\pi)=16$ kHz, respectively. Given the larger distance covered per drive period $T$ by edge modes in the anomalous regime, the corresponding data points are extracted from propagation over 20$T$ for a system size of $N_s = 1800$ and $L_y/L_x \approx2.7 $, whereas the same system size allows for a propagation up to 40$T$ without reaching the corner of the finite lattice in the Haldane regime.\\

At an intermediate modulation frequency of $\omega/(2\pi)=11.75$ kHz in the Haldane regime, the choice of initial state $| \varphi_0 \rangle$, tailored to overlap either with the wavefunction of edge states in the $\pi$-gap of the anomalous regime or in the $0$-gap of the Haldane regime, results in a vanishing or finite $\Delta \overline{y}$  in the absence of disorder. For non-vanishing disorder, if $|\varphi_0^{(\textrm{III})}\rangle$, the initial state shown in the left panel Fig.~\ref{fig:wavefunctions}c, is chosen, edge dynamics in the Haldane regime qualitatively reproduces the behavior at $\omega/(2\pi)=16$ kHz, away from the phase boundary, as displayed in \ref{fig:reply_fig}a by the numerical data in light red. Conversely, the re-entrant behavior of chiral motion observed experimentally (Fig.~\ref{fig:Fig4}c in the main text, dark red curve) is best reproduced by an initial state characterized by good overlap with edge states in the $\pi$-gap of the anomalous regime and vanishing overlap with edge states in $0$-gap of the Haldane regime. As extensively discussed above, a suitable choice to this end is $|\varphi_0^{(\textrm{II})} \rangle$, the wavefunction illustrated in the left panel of Fig.~\ref{fig:wavefunctions}b, resulting in the dark red curve in Fig.~\ref{fig:reply_fig}a, in good qualitative agreement with the experimental data for release from a shallow tweezer.

\section{Edge state signal near the phase transition}
\label{sec:1175}

To complement Fig.~4c, we show real-space images of the differential edge signal at $\omega/(2\pi) = \SI{11.75}{\kilo\hertz}$ for both the shallow as well as the deep tweezer configuration at selected disorder strengths (Fig.~\ref{fig:1175}). This comparison provides a direct visualization of the effect of disorder on the edge-state dynamics close to the phase transition between the anomalous Floquet topological and the Haldane regime.

\end{document}